\documentclass[12pt]{iopart}

\usepackage{graphicx}
\usepackage{subfig}
\usepackage{epsfig}
\usepackage{iopams}
\newcommand{\p}{\partial}

\newcommand{\wh}[1]{\widehat{#1}}

\newcommand{\vek}[1]{\mathbf{#1}}
\newcommand{\abs}[1]{\left|#1\right|}
\newcommand{\spar}{\shortparallel} 
\newcommand{\nablan}{\nabla_{\perp}}
\newcommand{\nablap}{\nabla_{\spar}}

\begin{document}
\title[]{Isotope effect on gyro-fluid edge turbulence and zonal flows}
\author{O~H~H~Meyer$^{1,2}$ and A~Kendl$^{1}$}
\address{$^1$Institute for Ion Physics and Applied Physics, University of Innsbruck, Association Euratom-ÖAW,
Technikerstr. 25, 6020 Innsbruck, Austria}
\address{$^2$Department of Physics and Technology, UiT - The Arctic University of Norway,
9037 Troms\o, Norway}
\ead{ole.meyer@uibk.ac.at}
\begin{abstract}
The role of ion polarisation and finite Larmor radius on the isotope effect
on turbulent tokamak edge transport and flows is investigated by means of
local electromagnetic multi-species gyro-fluid computations.
Transport is found to be reduced with the effective plasma mass for protium,
deuterium and tritium mixtures. This isotope effect is found for both cold and
warm ion models, but significant influence of finite Larmor radius and polarisation effects
are identified. Sheared flow reduction of transport through self generated
turbulent zonal flows and geodesic acoustic modes in the present model
(not including neoclassical flows) is found to play only a minor role on
regulating isotopically improved confinement. 
\end{abstract}
	
\submitto{\PPCF}
\maketitle
\section{Introduction}

The isotope effect in tokamak plasmas refers to experimentally measured improved
confinement properties with increasing ion mass $m_i$, as observed from
hydrogen (H) to deuterium (D) and tritium (T) plasmas \cite{bessenrodt93,hawryluk98}. 
While most present tokamak experiments for practical reasons run with
deuterium, actual fusion plasmas like in ITER will consist of the D-T plasma 
fuel mixed with the Helium ash. 

Whether specific theory based transport scaling models could, 
for certain underlying instabilities e.g. driven by trapped electron modes (TEM), 
roughly account for the experimental scalings has been controversial \cite{tokar04,waltz04,tokar04b}.  
In any case the observed improvement in confinement is apparently inconsistent
with primitive mixing length approximations of the turbulent gyro-Bohm cross-field
diffusivities $\chi \sim \rho_i \sim \sqrt{m_i}$, where
$\rho_i \sim \sqrt{m_i} $ is the ion gyro-radius.

A region of particular interest for transport in tokamaks is the plasma edge
around the last closed magnetic flux surface. Turbulence and transport in the edge
pedestal region crucially determine the low-to-high confinement transition,
the pedestal width, and further transport of particles and heat across the
scrape-off layer onto plasma facing components.

The isotope effect on computations of three-dimensional tokamak edge turbulence has been studied
initially with a drift-Alfv\'en fluid model, where the turbulent
particle and heat transport have been found to be only weakly dependent on ion mass \cite{scott97}. 
This was attributed to two competing mechanisms: a higher ion mass leads to more
adiabatic electron dynamics \cite{scott92} and consequently a weaker
radial transport (for a collisional coupling coefficient $C \sim 1/m_i$). This effect more or
less cancels the basic gyro-Bohm transport which is proportional to $\sqrt{m_i}$. 
The sensitivity was found to be larger for collisional electrostatic drift
wave transport compared to electromagnetic drift-Alfv\'en transport \cite{scott97}.

Initial computations on the isotope effect on ion temperature gradient (ITG)
core turbulence revealed a trend for favourable isotope scaling for the ion
thermal diffusivity, which had partly been attributed to the ITG growth rate
decreasing with isotope mass \cite{dong94,lee97}.

Recent experiments \cite{xu13}, theoretical work \cite{hahm13} and gyro-kinetic
computations of ITG and TEM turbulence \cite{pusztai11,bustos15} have
investigated the isotope effect on turbulent and/or neoclassical zonal flows, 
and on geodesic acoustic modes (GAMs) \cite{gurchenko16}. 
Increasing isotope mass has been found to lead to stronger zonal flows and GAMs, which
in turn reduce the radial turbulent particle transport magnitude. 

In this work we specifically focus on the influences of ion polarisation and ion temperature through finite Larmor radius (FLR) effects on
drift-Alfv\'en edge turbulence, zonal flows and GAMs
by means of three-dimensional isothermal gyro-fluid multi-species simulations for
various isotopic compositions.  

We introduce the model and numerical code employed in \Sref{s2}.
The mass dependence of our model is discussed in \Sref{s3}.
Results form numerical simulations are presented in \Sref{s4}, where we start with
a general observation of the isotope effect which is then studied more
systematically by considering various parts of the model independently. 
Conclusions and an outlook are presented in \Sref{s5}.

\section{Local isothermal gyro-fluid model and code} \label{s2}

In the local (delta-f) isothermal multi-species gyro-fluid model, the
normalised equations for the fluctuating gyro-center densities $n_s$ are \cite{scott05b}
\begin{equation} \label{density}
\frac{\rmd_{s} n_s}{\rmd t} = - g_{s} \frac{\p \phi}{\p y} - \nablap u_{s \spar} + \mathcal{K} \left( \phi_{s} + \tau_{s} n_{s} \right),  
\end{equation}
where the index $s$ denotes the species with $s \in (\mathrm{e}, i, j)$ for the main plasma
components (electrons with index $\mathrm{e}$ and main ions with index $i$) plus one or more
additional ion species with index $j$ (for simulations of mixtures of e.g. $i=$ deuterium
and $j=$ tritium).

The parallel velocities $u_{s \spar}$ and the vector potential $A_{\spar}$ evolve
according to \cite{scott05b}
\begin{eqnarray} \label{parvel}
\wh{\beta} \frac{\p A_{\spar}}{\p t} &+ \epsilon_{s} \frac{\rmd_s u_{s \spar}}{\rmd
  t} = - \nablap \left( \phi_{s} + \tau_{s} n_{s} \right) + 2 \epsilon_{s} \tau_{s} \mathcal{K} \left( u_{s \spar} \right) + \wh{\beta}
\tau_{s} g_{s} \frac{\p A_{\spar}}{\p y} - C J_{\spar}, 
\end{eqnarray}
and are closed by the gyro-fluid polarisation equation
\begin{equation} \label{poleq}
\sum_{s} a_{s} \left[ \Gamma_{1} n_s + \frac{\Gamma_{0} - 1}{\tau_{s}} \phi \right] = 0,
\end{equation}
and Ampere's law
\begin{equation}
- \nablan^{2} A_{\spar} = J_{\spar} = \sum_{s} a_s u_{s \spar}.
\end{equation}
The gyro-screened electrostatic potential acting on the ions is given by
\begin{equation*}
\phi_{s} = \Gamma_{1} \left( \rho_{s}^{2} k_{\perp}^ {2}  \right) \wh{\phi}_{\vek{k}},
\end{equation*}
where $\wh{\phi}_{\vek{k}}$ are the Fourier coefficients of the electrostatic potential.
The gyro-average operators $\Gamma_{0} (b) $ and $\Gamma_{1} (b) = \Gamma_{0}^{1/2} (b)$ correspond to
multiplication of Fourier coefficients by $I_{0}(b) e^{-b}$ and 
$I_{0}(b/2) e^{- b/2}$, respectively, where $I_{0}$ is the modified Bessel
function of zero'th order and $b = \rho_{s}^{2} k_{\perp}^ {2}$. 
We here use approximate Pad\'{e} forms with $\Gamma_{0} (b) \approx (1 + b)^{-1}$ and $\Gamma_{1} (b) \approx (1 + b/2)^{-1}$ \cite{dorland93}.

The perpendicular $\vek{E} \times \vek{B}$ advective and the parallel derivative operators
for species $s$ are given by 
\begin{equation*}
\frac{\rmd_{s} }{\rmd t} = \frac{\p }{\p t} + \left\{ \phi_{s}, ~ \right\},
\end{equation*}
\begin{equation*}
\nablap  = \frac{\p }{\p z} - \wh{\beta} \left\{ A_{\spar}, ~ \right\},
\end{equation*}
where we have introduced the Poisson bracket as
\begin{equation*}
\left\{ f, g \right\} = \left( \frac{\p f}{\p x} \frac{\p g}{\p y} - \frac{\p f}{\p y} \frac{\p g}{\p x}  \right).
\end{equation*}

In local three-dimensional flux-tube co-ordinates $\{x,y,z\}$, $x$ is a (radial) flux-surface
label, $y$ is a (perpendicular) field-line label and $z$ is the position along
the magnetic field-line.
In circular toroidal geometry with major radius $R$, the curvature operator is given by
\begin{equation*}
\mathcal{K} = \omega_{B} \left( \sin z \; \frac{\p}{\p x} + \cos z \; \frac{\p}{\p y} \right),
\end{equation*}
where $\omega_{B} = 2 L_{\perp} / R$,
and the perpendicular Laplacian is given by
\begin{equation*}
\nablan^{2} = \left( \frac{\p^{2}}{\p x^{2}} + \frac{\p^{2}}{\p y^{2}} \right).
\end{equation*}
Flux surface shaping effects in more general tokamak or stellarator geometry
\cite{kendl06,kendl03} are here neglected for simplicity. 

Spatial scales are normalised by the
drift scale $\rho_0 = \sqrt{T_\rme m_{i0}}/\rme B$, where $T_\rme$ is a reference
electron temperature, $B$ is the reference magnetic field strength and $m_{i0}$ is a reference ion mass, for which we use the mass of deuterium $m_{i0} = m_\mathrm{D}$.
The temporal scale is normalized by $c_0 / L_{\perp}$, where $c_0 = \sqrt{T_\rme/m_{i0}}$,
and $L_{\perp}$ is the generalized profile gradient scale length. 

The main species dependent parameters are
\begin{eqnarray*}
a_{s} = \frac{Z_s n_{s0}}{n_{\rme 0}} , \quad \tau_{s} = \frac{T_{s}}{Z_s T_{\rme}}, \quad
\mu_{s} = \frac{m_{s}}{Z_s m_{i0}}, \\
\rho_{s}^{2} = \mu_{s}\tau_{s} \rho_{0}^{2}, \quad \epsilon_{s} = \mu_{s} \left( \frac{q R}{L_{\perp}} \right)^{2},
\end{eqnarray*}
setting the relative concentrations, temperatures, mass ratios and FLR scales
of the respective species. $Z_s$ is the charge state of the species $s$ with mass $m_s$ and temperature $T_s$.

The gradient scale lengths for all species,
$g_{s} = \abs{\p \ln n_{s0} / \p x } = 1 / L_{n s}$ where $n_{s0}(x)$ is the respective
background density, satisfy the electrically neutral equilibrium condition $\sum_s a_{s} g_{s} = 0$. 

The plasma beta parameter
\begin{equation*}
\wh{\beta} = \frac{4 \pi p_{\rme}}{B^{2}} \left( \frac{q R}{L_{\perp}} \right)^{2},
\end{equation*}
controls the shear-Alfv\'{e}n activity, and
\begin{equation*}
C = 0.51 \frac{\nu_{\rme} L_{\perp}}{c_{0}} \frac{m_\rme}{m_{i 0}} \left( \frac{q R}{L_{\perp}} \right)^{2}, 
\end{equation*}
mediates the collisional parallel electron response for $Z=1$ charged hydrogen
isotopes.
The present computational study is based on the isothermal electromagnetic
multi-species gyro-fluid code \cite{scott03,scott05b}. This code features a globally consistent geometry \cite{scott98} with a shifted metric
treatment of the coordinates \cite{scott01}.  
The theoretical and computational basis of plasma edge turbulence models
and simulations is for example reviewed in \cite{scott07}.

We use an Arakawa-Karniadakis numerical scheme for the computations \cite{arakawa66,karniadakis91,naulin03}.  
The present multi-species code \cite{kendl14} has been cross-verified in the
local cold-ion limit with the tokamak edge turbulence standard benchmark case
of Falchetto \etal~\cite{falchetto08}, and with the results of finite Larmor
radius SOL blob simulations of Madsen \etal~\cite{madsen11}.

\section{Ion isotope mass dependence in the multi-species gyro-fluid model} \label{s3}

The isotope masses enter the multi-species gyro-fluid equations in several terms.
The term with the advective derivative of the parallel velocity in (\ref{parvel}) is
proportional to $\mu_s$ via $\epsilon_s$. 
The different ion masses also enter via the gyro-radius $\rho_s = \sqrt{\mu_s \tau_s} \; \rho_0$. This sets the magnitude 
of gyro-averaging operators depending on $b = \rho_s^2 k_{\perp}^2 = -\mu_s
\tau_s \nabla_{\perp}^2$.
Explicitly, mass dependent effects in the polarisation equation (\ref{poleq}) can be seen when expanding
$\Gamma_0 -1 = (1+b)^{-1}-1 = - b/(1+b) = \mu_s \tau_s \nabla_{\perp}^2 \Gamma_0$. 
For cold ions the polarisation equation thus reduces to
\begin{equation}
\sum_{s} a_{s} \left[ n_s + \mu_s \nabla_{\perp}^2 \phi \right] = 0,
\end{equation}
or
\begin{equation}
\mu_{\mathrm{eff}} \nabla_{\perp}^2 \phi = - \sum_{s}  a_{s} n_s, 
\label{pol-reduced}
\end{equation}
which depends on the effective mass $\mu_{\mathrm{eff}} = \sum_s a_{s} \mu_{s}$. 
For a single-ion plasma $\mu_{\mathrm{eff}} = a_{e} \mu_{e} + a_{i} \mu_{i}
\approx a_{i} \mu_{i} = 1$, when this ion mass is taken as the reference mass. 
In a multi-ion-species plasma the relative ion masses and concentrations
determine the effective polarisation through $\mu_{\mathrm{eff}}$.

For illustration of the isoptope effect on polarisation we construct a
multi-species quasi-2D gyro-fluid model, which corresponds to the 
classic Hasegawa-Wakatani (HW) model \cite{hasegawa83}. 
Here we assume a constant magnetic field, ignore curvature and electromagnetic effects and consider cold ions.
The parallel response then reduces to $C J_{\spar} = \nabla_{\spar} (n_e - \phi)$.
The multi-species gyro-fluid density equations in this limit are
\begin{equation} \label{dens-reduced}
\frac{\rmd}{\rmd t} n_s= - g_{s} \frac{\p \phi}{\p y} - \nablap u_{s  \spar}.
\end{equation}
We multiply (\ref{dens-reduced}) each with $a_s$ and sum up all equations:
\begin{eqnarray*} 
\frac{\rmd}{\rmd t} \left( \sum_s a_s n_s \right) = - \left( \sum_s a_s
g_{s} \right) \frac{\p \phi}{\p y} - \left( \sum_s a_s \nablap u_{s\spar} \right).
\end{eqnarray*}
Using $\sum_s a_{s} g_{s} = 0$, inserting (\ref{pol-reduced}) and
neglecting parallel ion dynamics yields ($u_{s \spar} = 0$)
\begin{equation*} 
\mu_{\mathrm{eff}} \frac{\rmd}{\rmd t} \nabla_{\perp}^2 \phi \approx  \nablap (a_\rme u_{\rme  \spar} ) \approx  \nablap J_{\spar}.
\end{equation*}
Defining $D \equiv k_{\spar}^2 / C$ as HW dissipative coupling coefficient (where $C$ is
normalised to the reference ion mass) and the electric drift vorticity $\Omega = \nabla_{\perp}^2 \phi$, the reduced model equations read: 
\numparts 
\begin{eqnarray}
\frac{\rmd}{\rmd t} n_\rme = D (\phi - n_\rme) - g_{\rme} \frac{\p \phi}{\p y}, \label{HWga}\\
\mu_{\mathrm{eff}} \frac{\rmd}{\rmd t} \Omega = D (\phi - n_\rme) \label{hwmult2}. \label{HWgb}
\end{eqnarray}
\endnumparts
The remaining gyro-fluids are passively advected while the potential $\phi$ can be determined from $\Omega$.
The mass of the secondary ion species thus enters via the effective
polarisation mass in the vorticity equation: higher effective mass reduces the
perpendicular inertial response and weakens the turbulent vorticity drive
maintained by  parallel dissipative coupling. 
The perpendicular wave number spectrum is rescaled by 
$ k_{\perp}^2 \rightarrow \mu_{\mathrm{eff}} k_{\perp}^2 \equiv K^2$, as 
$\wh{\Omega}_{\vek{k}} \rightarrow \wh{\Omega}_{\vek{k} {\mathrm{eff}}} = \mu_{\mathrm{eff}} \wh{\Omega}_{\vek{k}} = - K^2
\wh{\phi}_{\vek{k}}$, adapting to the drift scale given by the effective mass.

This can be illustrated by linear analysis of the multi-species HW equations,
which results in a real drift wave frequency of 
$\omega = g k_y/(1+K^2)$ and an approximate growth rate (in the
weakly nonadiabatic limit) $\gamma \approx (\omega / \omega_d)^2$ 
with $\omega_d^2 = D(1+K^2)/K^2$. The drift wave frequency is reduced and the growth rate enhanced for increasing effective mass.

Similar to this inertial isotope mass effect on polarisation through the drift
scale, we can expect gyro-screening and gyro-averaging FLR effects for warm
ions through the gyro-scales. 
 
In summary, the ion isotope mass enters into the isothermal 3-dimensional
gyro-fluid equations in three ways:
\noindent ($\rmi$) Re-scaling of the effective drift scale 
$\rho_0 \rightarrow \sqrt{\mu_{\mathrm{eff}}} \; \rho_0$ in the ion polarisation
entering through the term $(\Gamma_0 -1) / \tau_s$ in (\ref{poleq}).
\noindent ($\rmi \rmi$) Re-scaling of the ion gyro-scales 
$\rho_s = \sqrt{\mu_s \tau_s} \; \rho_0$ in the gyro-screening and
gyro-averaging FLR operators for finite ion temperatures $\tau_s > 0$. 
\noindent ($\rmi \rmi \rmi$) Parallel ion response scaled by $\epsilon_s \sim \mu_s$ in (\ref{parvel}).

As for any (of arbitrary multiple) ion species $|\mu_\rme| \ll |\mu_i|$ the
adiabatic and electromagnetic response is mainly carried by the electrons,
we do not expect the third effect to be of major relevance.

In the following we numerically study the relative importance of isotope
polarisation and FLR effects with the complete 3-dimensional isothermal
gyro-fluid model in order to identify their roles on turbulence, transport and flows.

\section{Computational results} \label{s4}

Nominal reference simulation, and roughly corresponding physical, parameters are listed in Table \ref{tabtwo}.
\begin{table}
\caption{\label{tabtwo}Numerical and physical parameters.} 
\begin{indented}
\lineup
\item[]\begin{tabular}{@{}*{7}{l}}   
\br
\m$\wh{\beta}$&\m$\wh{s}$&\m$\omega_{B}$&\m$\tau_s$&\m$C$&\m$\left( q R / L_{\perp} \right)^{2}$\\
\m$1$&\m$1$&\m$0.05$&\m$1$&\m$5$&\m$\018,500$\\
\mr
\m$n_\rme ~ [\mathrm{cm}^{-3}]$&\m$T_\rme ~ [\mathrm{\rme V}]$&\m$B ~ [\mathrm{T}]$&\m$R ~ [\mathrm{cm}]$&\m$L_{\perp} ~ [\mathrm{cm}]$&\m$\0q$\\
\m$\03 \times 10^{13} $&\m$\0\070$&\m$\02.5$&\m$\0165$&\m$\04.25$&\m$3.5$\\
\br
\end{tabular}
\end{indented}
\end{table}
To neglect electromagnetic effects, we also perform simulations with $\wh{\beta} = 0$. 
All simulations are initialised with the same turbulent bath given by the sum
over multiple small-amplitude sinusoidal modes in density fluctuations. 
All simulations are carried deeply into the non-linearly satured state which,
depending on plasma composition, typically sets in at $t \sim 200$ normalized
time units. 
Statistics are sampled over at least $8,000$ time units in the turbulent
state. In the subsequent plots the standard deviation
of fluctuations around the mean of the respective time series is used as error bars. 
Numerical dissipation of the form $\nu_{\spar} \p_{zz}^{2} - \nu_{\perp} \nabla_{\perp}^{4}$ with $\nu_{\spar} = 0.01$ and $\nu_{\perp} = 0.02$ is added to each
advective derivative.

\subsection{Transport scaling with effective mass}

We consider two ion species with mass ratios $\mu_{i}, \mu_{j}$ and
concentrations $a_{i}, a_{j}$ and compare results of turbulence computations
with respect to the effective mass $\mu_{\mathrm{eff}} = a_{i} \mu_{i} + a_{j} \mu_{j}$ 
both for cold ions ($\tau_i = \tau_j = 0$) and warm ions ($\tau_i = \tau_j =1$).

\begin{figure} 
  \begin{center}
    \subfloat{\includegraphics[width=65mm]{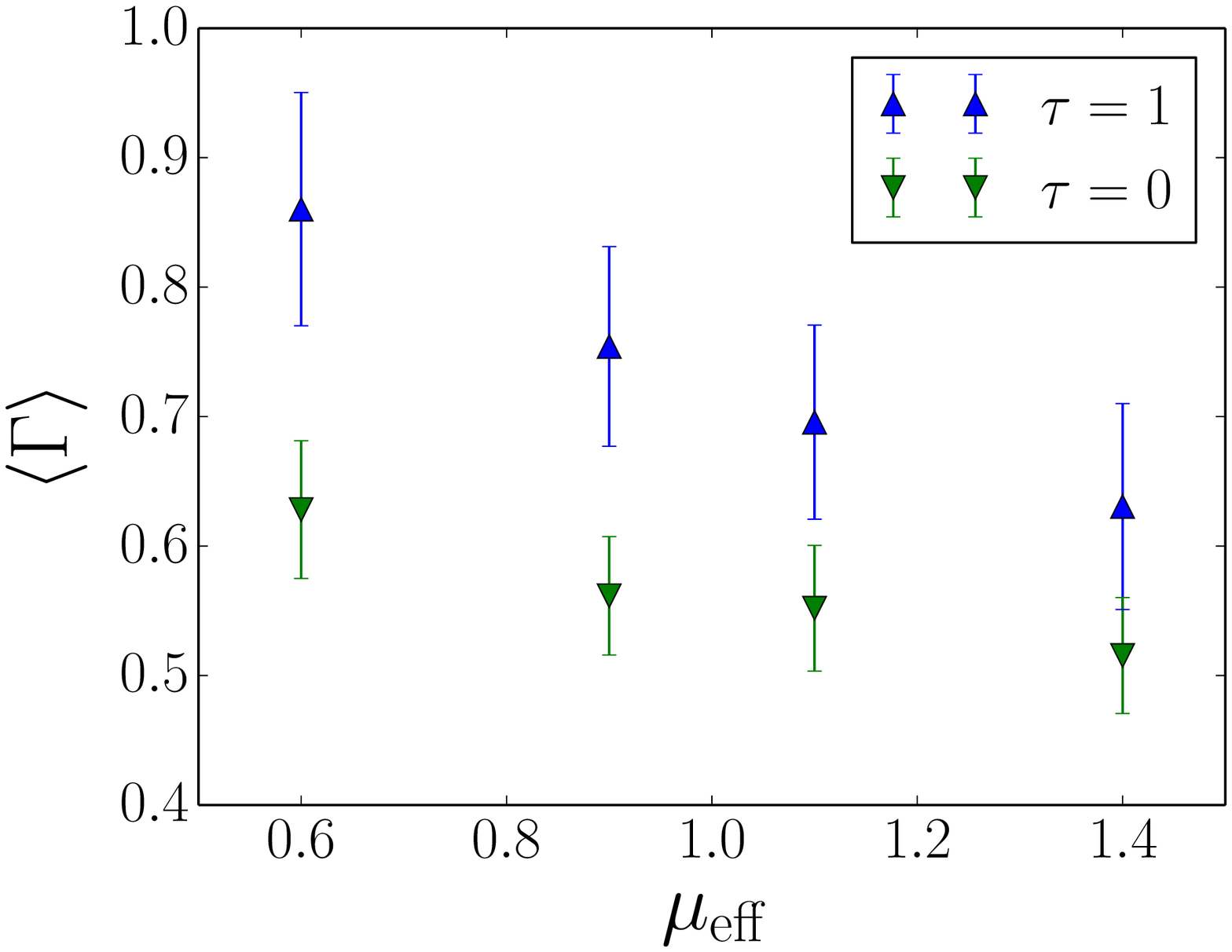}} \par
    \subfloat{\includegraphics[width=65mm]{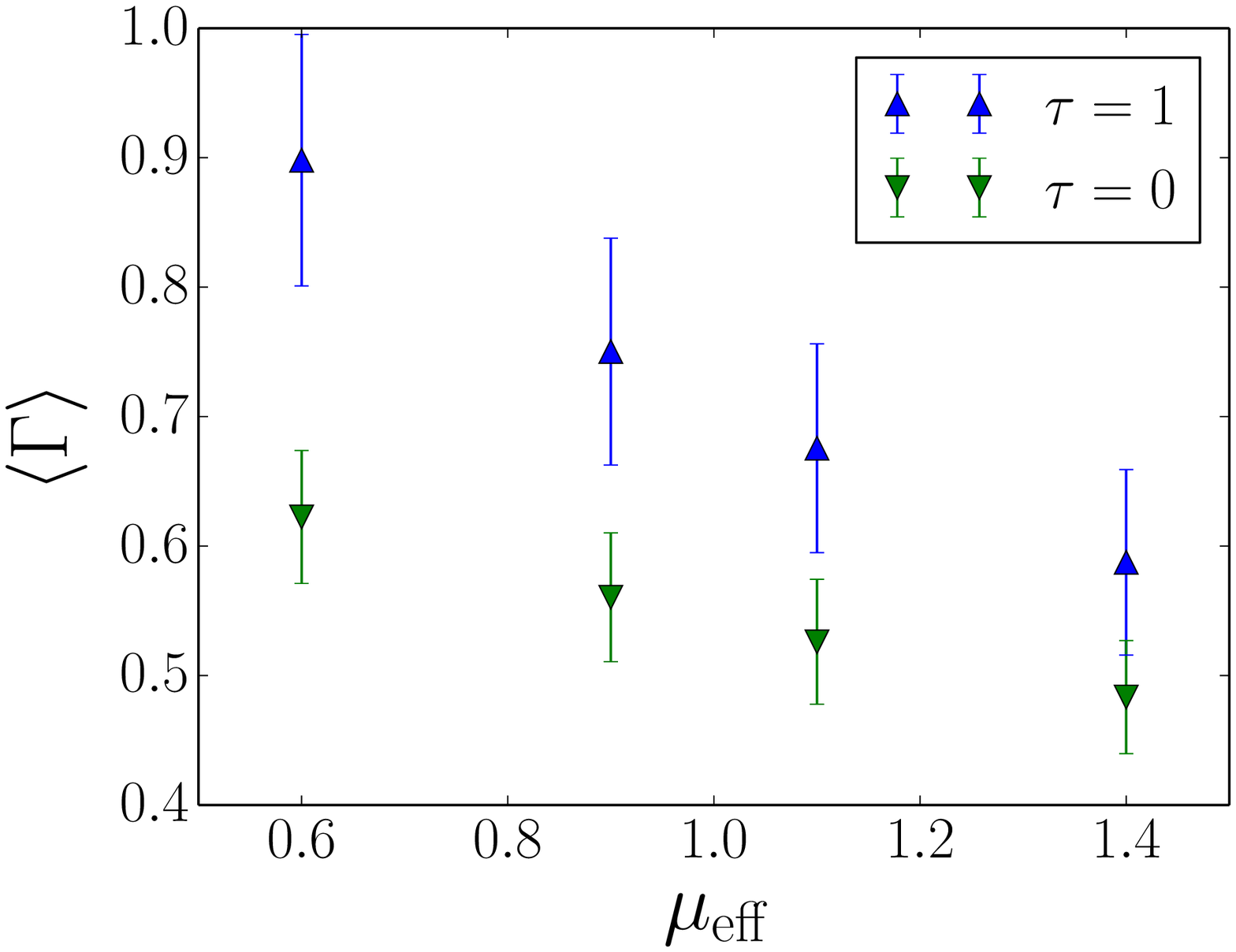}}
  \end{center}
\caption{Average radial particle flux $\langle \Gamma \rangle$ for varying effective ion mass $\mu_{\mathrm{eff}}$,
with and without FLR effects for warm and cold ions, respectively. Electrostatic $\wh{\beta} = 0$ (top) and electromagnetic $\wh{\beta} = 1$ (bottom). }  
\label{concscan}
\end{figure}

\begin{figure} 
  \begin{center}
    \subfloat{\includegraphics[width=65mm]{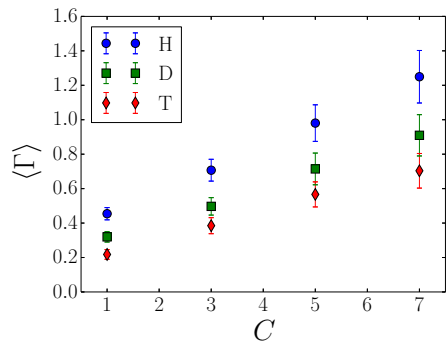}}
  \end{center}
\caption{Average radial particle flux $\langle \Gamma \rangle$ as a function of
  electron-ion collisionality parameter $C$ for pure H, D and T plasmas.} 
\label{cscan_gamma}
\end{figure}

The range of $\mu_{\mathrm{eff}} = 0.6$ to $\mu_{\mathrm{eff}} = 1.4$ includes 
protium-tritium-deuterium mixtures ranging from almost pure protium to almost pure tritium, see Table \ref{tabone}. 
\begin{table}
\caption{\label{tabone}Values of effective mass with corresponding plasma compositions.} 
\begin{indented}
\lineup
\item[]\begin{tabular}{@{}*{7}{l}}   
\br
\m$\mu_{\mathrm{eff}}$&\m$0.6$&\m$0.9$&\m$1.1$&\m$1.4$\\
\mr
$\mathrm{H}:\mathrm{D}$&$80:20$  &$20:80$&\m---&\m---&\\
$\mathrm{D}:\mathrm{T}$&\m---&\m---&$80:20$&$20:80$\\
\br
\end{tabular}
\end{indented}
\end{table}
In Figure~\ref{concscan} the average radial particle flux $\langle \Gamma \rangle = \langle
{n_\rme} v_x \rangle$, averaged both spatially over the simulation domain and
temporally, is shown as a function of the
effective mass $\mu_{\mathrm{eff}}$ in a plasma with two ion species. 
Normalisation is with respect to the reference ion mass, in our case deuterium.
The transport clearly decreases with increasing effective mass for both, cold and warm ions. 
Increased transport levels for warm ions relative to the cold
ion case are caused by the additional ion diamagnetic contribution to the
curvature coupling term ${\cal K} (\tau_i n_i)$ in (\ref{density}).
Reduction of transport with effective mass is stronger for warm
ions, which points to a pronounced role of the isotope mass on gyro-screening
through FLR effects. Transport reduction for heavier plasmas shows little dependence on finite $\wh{\beta}$.
We now test that the isotope mass effect found above is actually robust with
variation of the collisionality parameter $C$ from $1$ to $7$. 
Results are shown in Figure~\ref{cscan_gamma}. 
Plasma compositions considered are close to pure cases, i.e. H
denotes a 99~\% protium plasma with 1~\% deuterium. D is pure deuterium and T
denotes 99~\% tritium with 1~\% deuterium. 
It can be seen that transport levels for each species increase with larger
collisionality $C$, which is consistent with standard benchmark cases \cite{falchetto08}. 
The isotope effect is present throughout this range of collisionalities,
which indicates that the collisional response, which mainly governs the parallel
electron dynamics for the present range of parameters is not the only and
major mechanism acting on transport reduction with respect to isotope mass.

\subsection{Mass effects on zonal flows and GAMs}

We next examine the role of zonal flows on the reduction of turbulent
transport with increasing effective mass.

\begin{figure} 
  \centering
  \begin{center}
    \subfloat{\includegraphics[width=65mm]{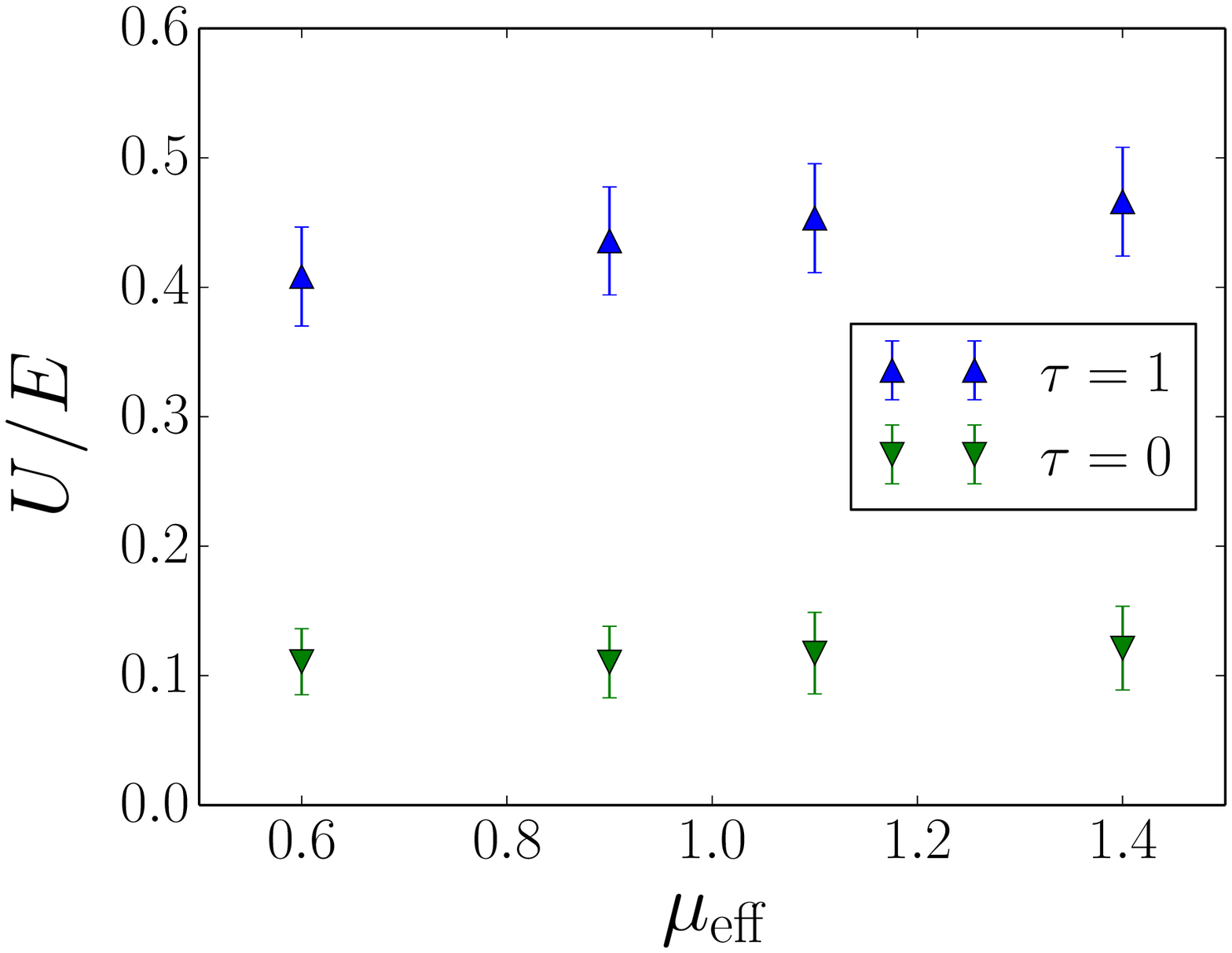}} \par
    \subfloat{\includegraphics[width=65mm]{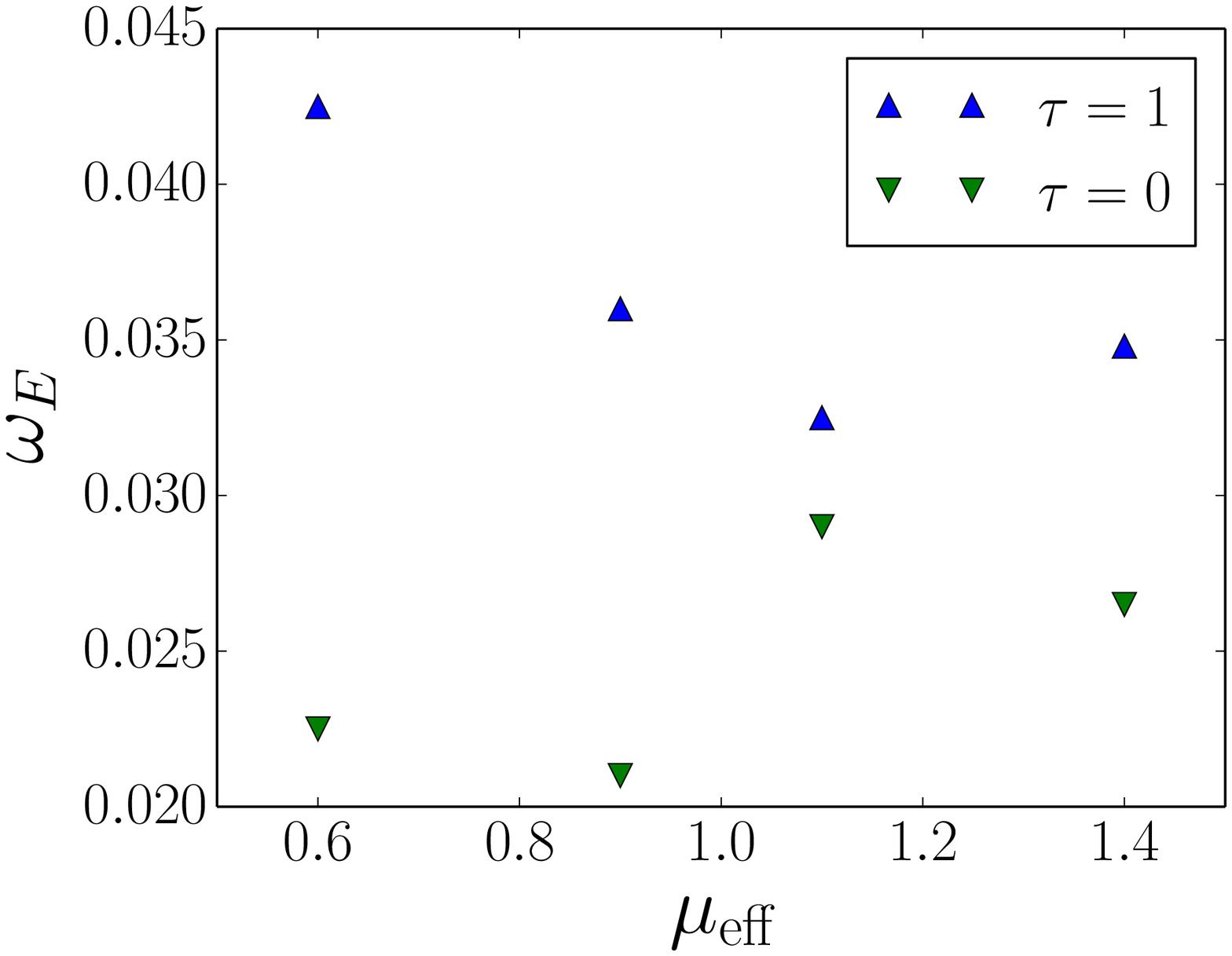}}
  \end{center}
\caption{Top: Relative zonal flow energy intensity $U/E$ as a function of effective
  ion mass $\mu_{\mathrm{eff}}$  for warm and cold ions. 
 Bottom: Zonal flow shearing rate $\omega_{E}$ for cold- and warm-ions.}  
\label{shearing}
\end{figure}

In Figure~\ref{shearing} (top) the ratio between zonal flow (ZF) energy $U = \sum_{s} a_{s} \int \mathrm{d}x \langle \phi_{s} \rangle \langle n_{s} \rangle / 2$, 
where $\langle \cdot \rangle = \int \mathrm{d}y \mathrm{d}z$ is the flux-surface average, and
total turbulent kinetic energy $E = \sum_{s} a_{s} \int \mathrm{d}x \rmd y \rmd z \phi_{s} n_{s} / 2$ as a measure for the zonal flow strength is
plotted as a function of the effective mass. 
While the ZF strength is generally larger for the warm ion case, both
cold and warm ion cases show little dependence of the ZF strength on the
effective mass: for $\tau_i=0$ the ZF energy ratio is nearly unaffected, and for $\tau_i=1$ it increases by approximately 5~$\%$
when varying from $\mu_{\mathrm{eff}} = 0.6$ to $1.4$. 

Another mechanism by which ZFs may reduce radial transport is by poloidal flow shear. Here, the shearing rate
$\omega_{E}$ is defined as the inverse auto-correlation time of $\p_{x}^2 \langle \phi \rangle (x_p)$, 
averaged over six distinct radial positions $x_p$. This is shown in Figure~\ref{shearing} (bottom), again for
cold and warm ions.
While the shearing rates on average are about 40~$\%$ stronger for the warm ion
case compared to cold ion case, they are again nearly unchanged (within
statistical uncertainties) for variations of the effective mass.

We have conducted analyses of a number of further statistical quantities as a
function of the effective mass. 

\begin{figure} 
  \begin{center}
    \subfloat{\includegraphics[width=65mm]{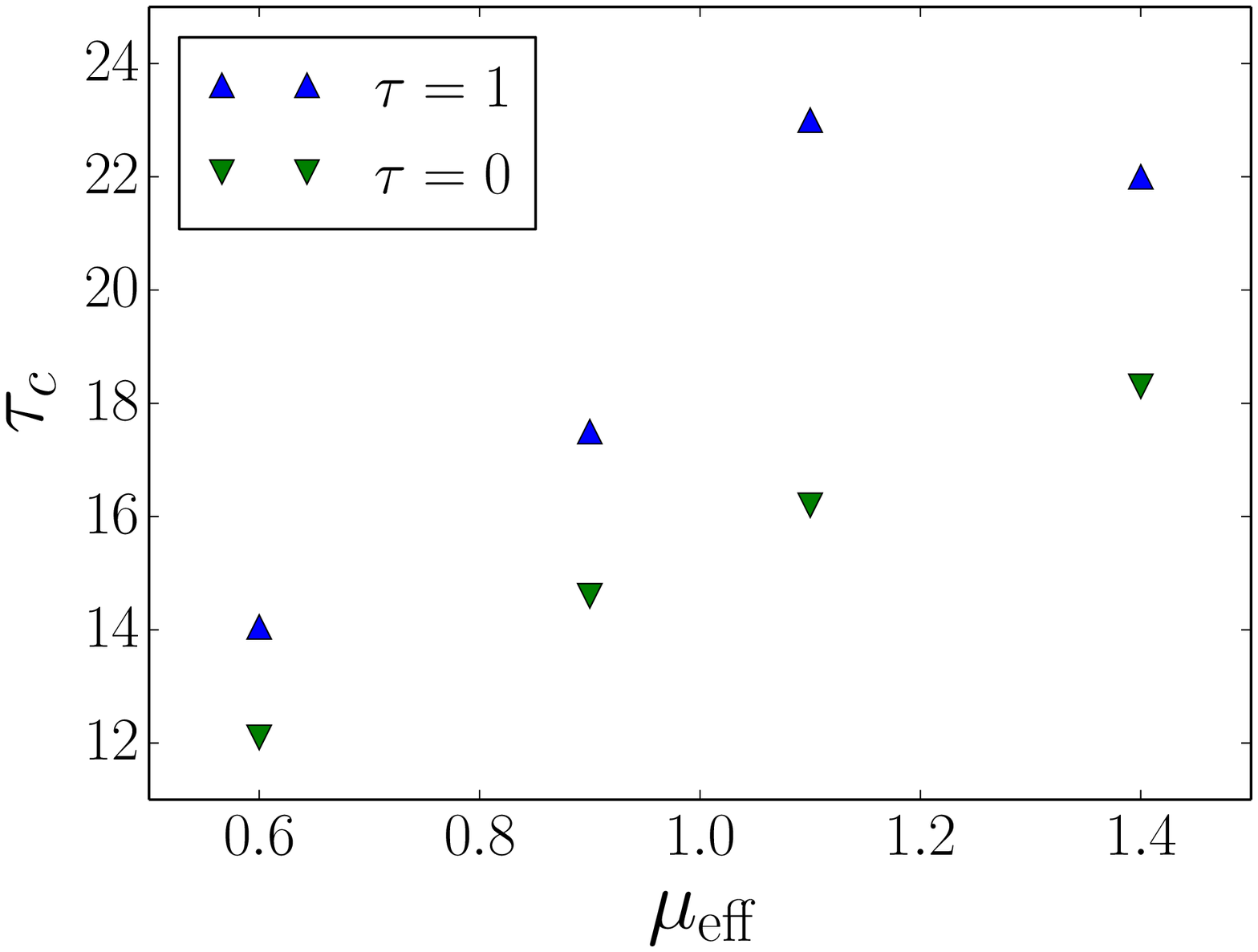}} \par
    \subfloat{\includegraphics[width=65mm]{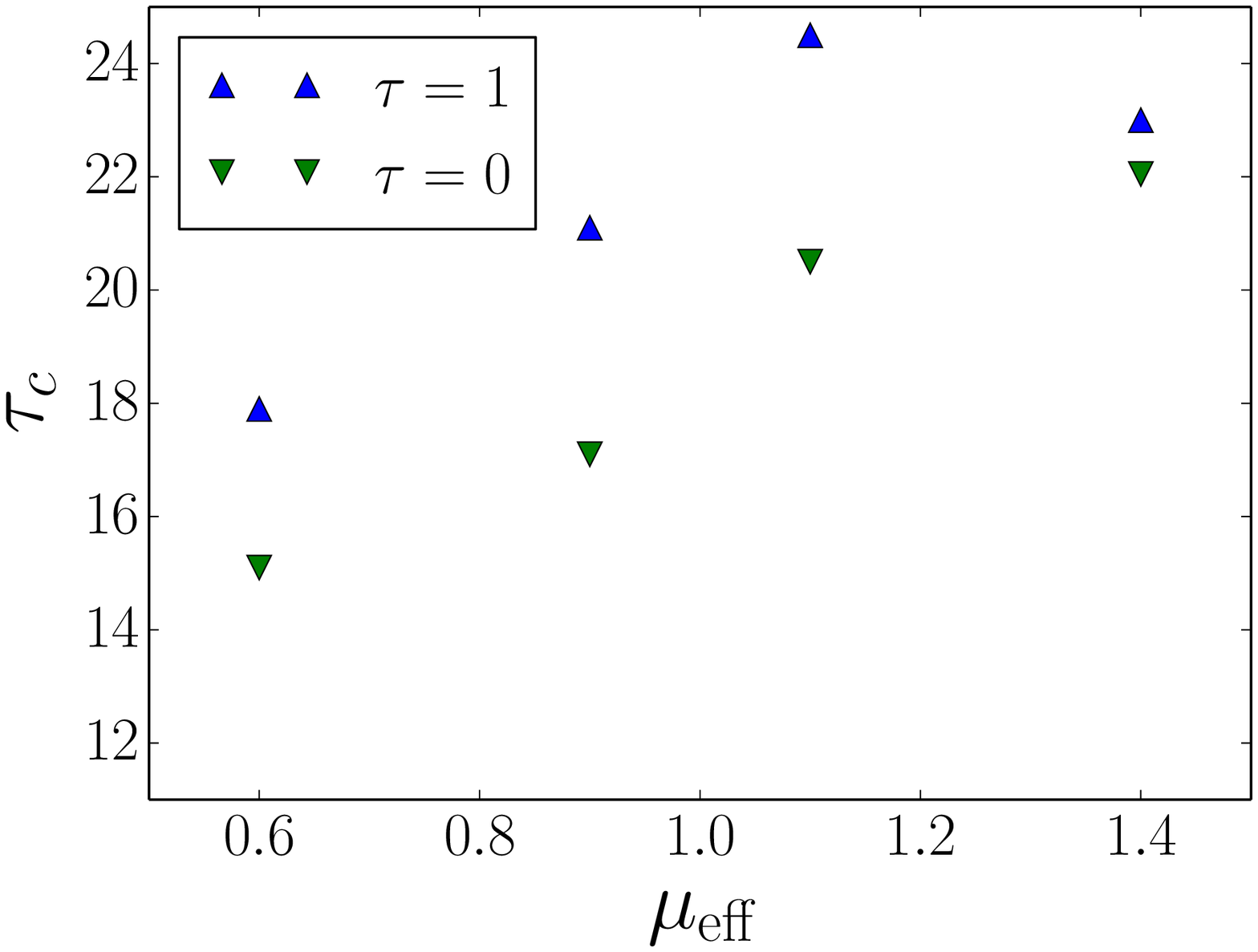}}
  \end{center}
\caption{Correlation times for density (top) and potential (bottom).}
\label{corrtimes}
\end{figure}

Characteristic time scales $\tau_c$ can be deduced from auto-correlation functions,
where state variables are sampled at each time step to produce time-series at $x_p$. 
Figure~\ref{corrtimes} shows, both for density (top) and potential (bottom)
auto-correlation times, the general trend of longer time scales
as the effective plasma mass increases.  
This trend is in approximate agreement with the shift of the drift wave frequency 
$\omega \sim 1/(1+\mu_{\mathrm{eff}})$ for the wave number scale $(\rho_0 k_{\perp})=1$,
and thus $\tau_c \sim 1+\mu_{\mathrm{eff}}$: the time scale increases by around
50~$\%$ when varying the effective mass from $\mu_{\mathrm{eff}}=0.6$ to $1.4$.

\begin{figure} 
  \begin{center}
    \subfloat{\includegraphics[width=65mm]{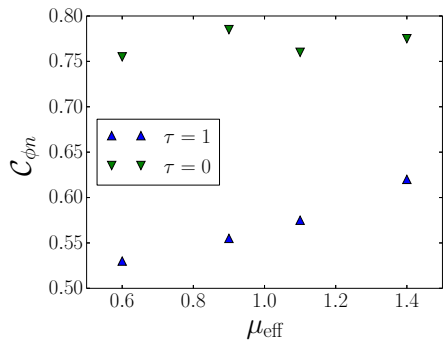}}
  \end{center}
\caption{Maximum cross-correlation between density and potential for cold and
  warm ions.} 
\label{corrmax}
\end{figure}

As shown in Figure~\ref{corrmax}, we also find an increase in the maximum of the
cross-correlation between density and potential fluctuations for warm ions,
but not for the cold ion case. 
Values stem from averages of correlation times and cross-correlations over two
time series at two distinct locations on the outboard midplane.

Typical values of the cross-correlation coefficient computed for the warm ion case are approximately a third weaker than in the cold ion case, but increase
with effective mass as opposed to the cold ion case.
From these trends we can infer that, first, the general reduction of cross-correlation
for warm ions is a result of the $\tau_i$ dependence in the diamagnetic curvature coupling,
which facilitates a trend towards a more interchange-like phase shift
distribution and thus weaker cross-correlation. 
Second, the increase of the cross correlation with effective mass for the warm
ions then can be attributed to warm ion dynamics, which by gyro-averaging counteract
the phase shift.

\begin{figure} 
  \begin{center}
    \subfloat{\includegraphics[width=65mm]{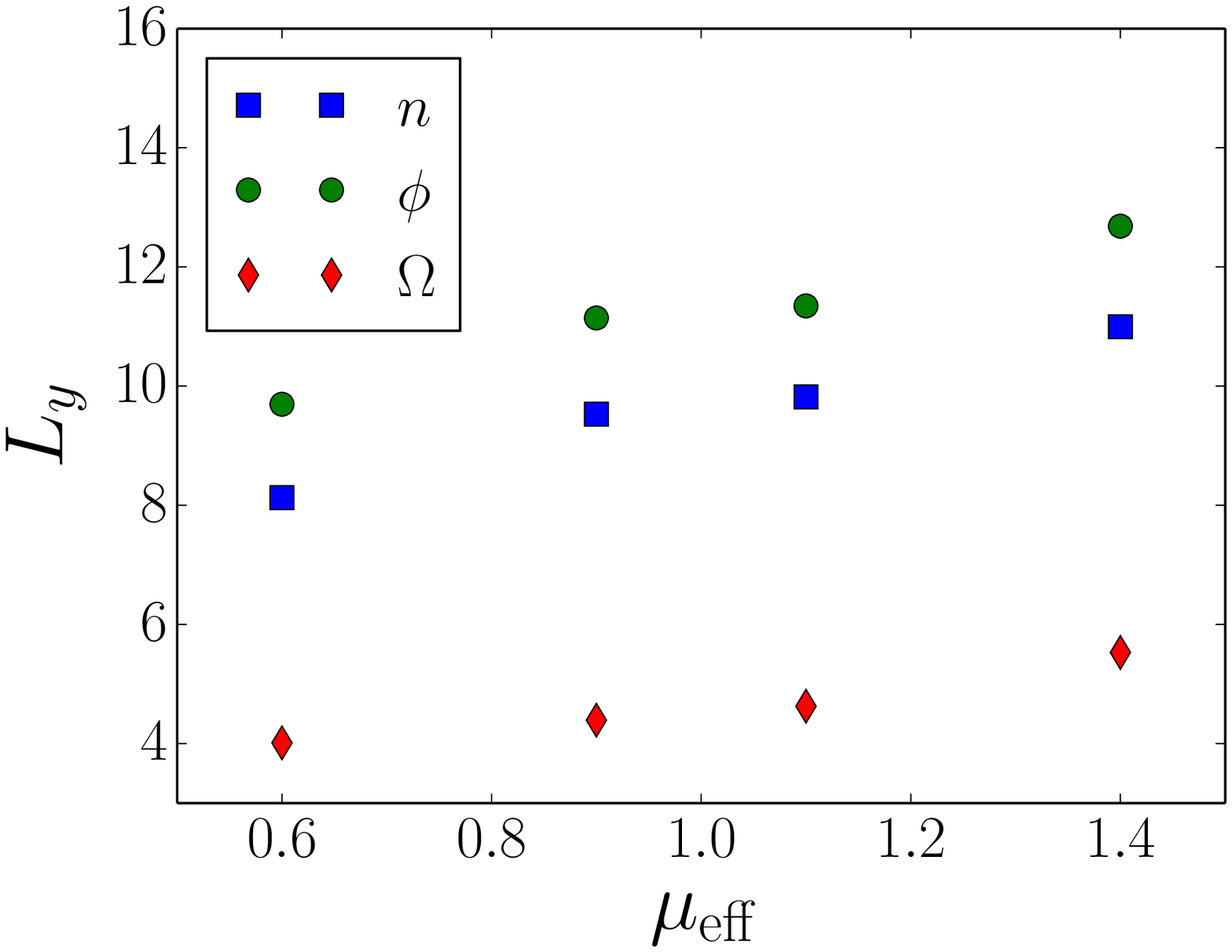}} \par
    \subfloat{\includegraphics[width=65mm]{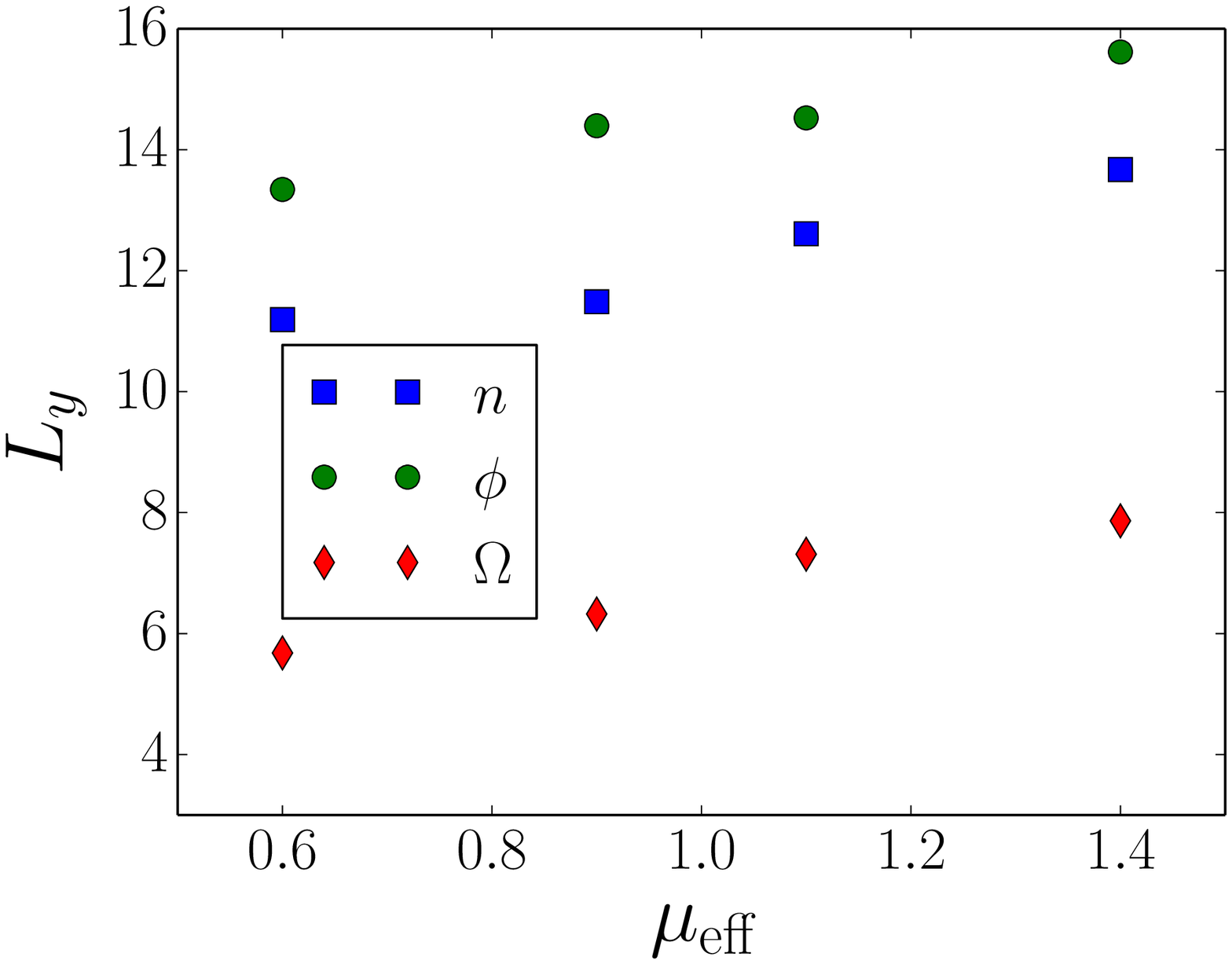}}
  \end{center}
\caption{Correlation lengths for cold- (top) and warm ions (bottom).}
\label{corrlengths}
\end{figure}

Figure~\ref{corrlengths} shows correlation lengths $L_y$ in the perpendicular $y$-direction of the density, electric potential, and vorticity fields. 
For all cases we find that $L_y$ increases by approximately 20~$\%$ when varying the effective
mass from $0.6$ to $1.4$.
As the scaling is similar for both cold and warm ions, we infer that this trend could be 
based on the drift-scaling (which is independent of $\tau_i$ in contrast to
direct FLR effects) by the effective mass acting on the polarisation.

On the other hand, an increase of perpendicular correlation lengths could also
be attributed to stronger zonal flows. 
Here, the results allow no unambiguous conclusion: while the relative ZF strength and the
shearing rate both only show weak dependence on the effective mass for warm ions and not at all for cold ions, the perpendicular
(``poloidal'') correlation length increases for both cold and warm ions by
around 20~$\%$.

In the context of the present (isothermal) gyro-fluid model, turbulence driven
zonal flows thus only play a rather minor role as cause for the overall isotope
effect on turbulent transport reduction.

Regarding the influence of isotope mass on GAMs,
we have found no dependence of the GAM intensity (quantified by the spectral power of $\langle \phi \rangle$ at the GAM frequency) in our isothermal gyro-fluid
simulations for the reference parameters.  

We observe that in regions of strong flow shear, i.e. close to the radial boundaries,  
the ZFs are strong and the GAM activity is weak, whereas the ZF activity is
weak in the center of the radial domain and GAMs are strong, consistent with
energetic exchange of flux-surface averaged kinetic flow energy between zonal
flows and GAMs. 

\subsection{Mass effects through finite Larmor radius and polarisation}

In order to separate the mass effects acting through FLR terms or the drift scaling
of polarisation, we set up a number of computations with artificially modified equations.

\begin{figure} 
  \begin{center}
    \subfloat{\includegraphics[width=65mm]{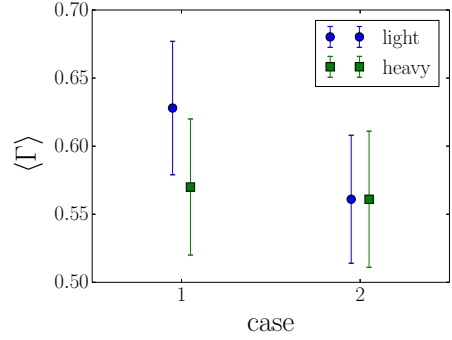}}
  \end{center}
\caption{Average radial particle flux $\langle \Gamma \rangle$ for standard model ($\rmi$) and modified model ($\rmi \rmi$).}
\label{transport_t0}
\end{figure}

First, we study the effect of mass on transport through polarisation by modifying the standard
cold ion case. Set one of simulations is run with the usual effective mass
defined by $\mu_{\mathrm{eff}} = a_i \mu_i + a_j \mu_j$ given by separate mass
ratios $\mu_i$ and $\mu_j$, here taken for hydrogen and deuterium masses, respectively.
Set two is run with artificially setting $\mu_{i} = \mu_{j}$ in the
effective mass entering in the polarisation.
We perform both sets of runs with ion concentrations $a_\mathrm{H}:a_\mathrm{D} = 80:20$ and $a_\mathrm{H}:a_\mathrm{D} = 20:80$,
denoted by "light" and "heavy", respectively.
The resulting transport rates are given in Figure~\ref{transport_t0} (data points are shifted slightly for better visibility).
We find that the modified mass dependence in the vorticity term of the
polarisation equation effectively annihilates the isotope effect, confirming
the above assumptions. This now is not surprising any more, since for cold
ions this term is the only mass-dependent term of relevance in the equations. 

Performing similar modifications in order to focus on specific FLR effects
only is a bit more tedious for the warm ion case, since both gyro-averaging
operators are mass dependent. 
We recall the warm ion gyro-screening term, $(\Gamma_{0,s} -1) / \tau_{s} = -
\mu_{s} k_{\perp}^2 / (1 + \mu_{s} \tau_{s} k_{\perp}^2) $, occuring in the
polarisation equation with the vorticity; and the gyro-averaging operator,  
$\Gamma_{1, s} = 1 / (1 + \mu_{s} \tau_{s} k_{\perp}^2)$, present in the
averaging of density as part of the polarisation equation and acting on the
potential in curvature and  parallel terms. The latter is most significantly
as part of the $\vek{E} \times \vek{B}$ advection.

\begin{figure} 
  \begin{center}
    \subfloat{\includegraphics[width=65mm]{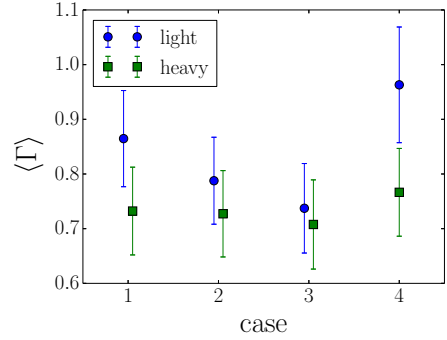}}
  \end{center}
\caption{Average radial particle flux $\langle \Gamma \rangle$ for the warm ion standard model 1 and cases 2--4 as
  described in the text.} 
\label{transport_t1}
\end{figure}

We introduce three modifications of the model to artificially enhance and
study the relative importance of each of these FLR effects. 
\begin{itemize}
\item Case 1 denotes the original model.

\item In case 2 we set $\Gamma_{0,i} = \Gamma_{0,j}$  in the ``vorticity'' term
without modifying the FLR operators elsewhere.   

\item Case 3 leaves $\Gamma_{0,s}$ unaltered, but sets $\Gamma_{1, i} = \Gamma_{1,j}$ 
on the gyro-fluid densities in the polarisation equation (and also in the
"vorticity-free" boundary condition). Other occurences of $\Gamma_{1,s}$ are  left unchanged. 

\item Finally, case 4 considers the impact of the gyro-averaged potential in the
advective terms by setting $\Gamma_{1,i} = \Gamma_{1,j}$ there without modifying it in the
polarisation equation and vorticity boundary conditions. 
\end{itemize}
Results for the particle transport for all cases are given in Figure~\ref{transport_t1}, with data points shifted slightly for improved visibility where necessary.  
The stronger the separation of the results between the ``light'' and ``heavy''
cases deviates from the standard case 1 in Figure~\ref{transport_t1}, the more
pronounced is the mass effect in the three different modified cases 2--4,
which highlight the relative importance in the FLR terms.

It can be seen that both mass-dependent FLR terms in the polarisation equation
contribute significantly to the isotope effect. The main contribution is
apparently caused by gyro-averaging on the gyro-fluid densities, whereas
modification of gyro-screened advective terms actually leads to enhancement of
the isotope effect.

\subsection{Mass effects on ion inertia}

\begin{figure} 
  \begin{center}
    \subfloat{\includegraphics[width=65mm]{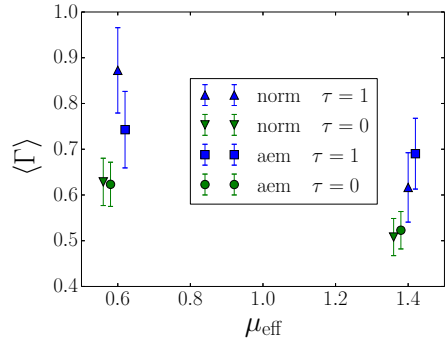}}
  \end{center}
\caption{Average radial particle flux $\langle \Gamma \rangle$ for adapted electron masses (aem) compared to runs of the
  full model (norm).} 
\label{elmass}
\end{figure}

We filter out inertia mass effects by introducing artifical electron-ion
mass ratios for single-ion (D) computations, similar as in Bustos \etal~\cite{bustos15}. 
The electron mass is adjusted such that the ion-electron mass ratio is the same as 
for the regular cases discussed above. 
In this case the gyro-operators $\Gamma_0$ and $\Gamma_1$ are evaluated at $b = \rho_0^2 k_{\perp}^2$, but the electron mass is varied
according to $\mu_{\rme} \rightarrow \mu_{\rme} / \mu_{\mathrm{eff}}$, such that an isotope mass scaling is introduced in the inertia only.
Results are shown in Figure~\ref{elmass}. 

For the cold ion case, there are no FLR effects and the only source for differences in
transport levels is due to the plasma inertia. 

We consider cases with $\mu_{\mathrm{eff}} = 0.6$ and $\mu_{\mathrm{eff}} = 1.4$
resulting from $\mathrm{H}:\mathrm{D} = 80:20$ and $\mathrm{H}:\mathrm{D} = 20:80$, respectively. In Figure~\ref{elmass} data points are shifted slightly around these values
for better visibility.

For cold ions, both the adapted electron mass model and the
reference model as expected yield comparable transport scaling, as expected. 

For warm ions, we find that
the relative inertia is only of rather minor importance compared to resolving the true gyro-radii. 
The isotope effect solely based on plasma inertia with identical Larmor radii 
is far weaker than the fully resolved case, indicating that the underlying
mechanism is maintained from both inertia and FLR effects. 

\subsection{Mass-adapted box and grid size test}

\begin{figure} 
  \begin{center}
    \subfloat{\includegraphics[width=65mm]{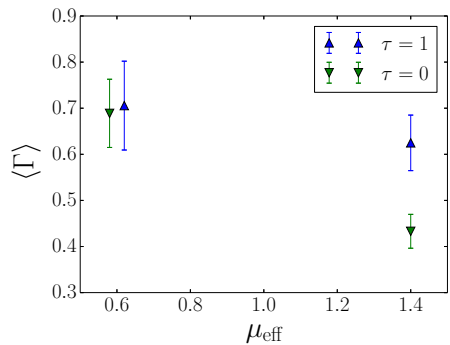}}
  \end{center}
\caption{Average radial particle flux $\langle \Gamma \rangle$ for adapted simulation domain according to effective plasma mass.}
\label{area}
\end{figure}

Since a species Larmor radius $\rho_s^2 = \mu_s \tau_s \rho_0^2 \sim \mu_{\mathrm{eff}}$
we adapt the perpendicular simulation box size to yield the same number of
gyro-radii per drift plane. 
This effectively increases the perpendicular resolution for $\mu_{\mathrm{eff}} < 1$ 
and decreases it for $\mu_{\mathrm{eff}} > 1$. 
Figure~\ref{area} shows results for $\mu_{\mathrm{eff}} = 0.6$ and $\mu_{\mathrm{eff}} = 1.4$ (data points are shifted where necessary for better visibility). 
These runs feature drift plane dimensions of $\{38 \times 154\} \rho_{0}$ and $\{90 \times 358\} \rho_{0}$ respectively. 
For all runs the number of nodes is $n_x \times n_y = 64 \times 256$. 

Resulting transport levels in Figure~\ref{area} can be compared with the
reference results labelled 'norm $\tau=0$' and 'norm $\tau=1$' in Figure~\ref{elmass}, which are obtained from the standard resolution $\{96 \times 256\} \rho_{0} \times 2 \pi$
on $64 \times 256 \times 16$ nodes. 

The mass-adapted resolution yields slightly higher transport levels for the
light plasma and slightly lower transport levels for the heavy plasma
(compared to the standard resolution). However, the isotope effect is still
clearly visible. Hence the results shown in the previous
subsections are not artificially modified by box size effects.

The box size variations have taken into account possible effects of the vortex
and zonal flow scales which fit into the computational domain.

Further computational or numerical artefacts could occur through the radial
(Dirichlet) boundary conditions and through insufficient spatial resolution. 
We have also analysed the dependence of such computational errors on the isotope
mass by means of the energy theorem. These results have confirmed that the
employed resolution and treatment of boundary conditions do not significantly modify
the overall isotope effects found in this work.

\section{Discussion and conclusions} \label{s5}

We have discussed ion isotope mass effects on electromagnetic gyro-fluid
tokamak edge turbulence and zonal flows. 
The simulations show reduced transport amplitudes in heavier plasmas 
in terms of the reduced radial particle flux. 

Zonal flow activity is found to be slightly enhanced with the isotope mass, but not to
an extent to be able to (fully) explain the observed transport reduction.
In the present model (isothermal, without mass dependent ion-ion collisions)
we do not find any significant influence of isotope mass on GAM amplitudes.
 
The isotope effect is stronger for warm ions ($\tau_i=1$) but persists also
for cold ion cases ($\tau_i=0$), which suggests that not FLR effects are the
main mechanism for transport reduction, but ion inertia effects in the
polarisation mediated by the mass-dependent drift scale.
This is also supported by artificially adapting the electron mass relative to
the ion mass to enhance pure ion inertia effects. 

In the presented simulations with finite collisionality, electromagnetic flutter
effects show no significant dependence on isotopes. However we note that in
our present gyro-fluid model implementation the parallel magnetic vector
potential is not subject to gyro-averaging operators, and thus electromagnetic
FLR effects are neglected \cite{scott05b}.

While the isotope effect is thus clearly present in the local isothermal gyro-fluid
model, we still find a weaker influence on transport reduction and zonal flow
enhancement compared to experimental results. 
This discrepancy could be caused by the neglect of thermal fluctuations and
ion temperature gradient (ITG) and trapped electron mode (TEM) turbulence in
the present model. Isotope mass effects on zonal flow damping via ion-ion
collisionality and on neoclassical flows can be studied with full thermal
models. In the future our delta-f gyro-fluid model is going to be extended to 6
moments including temperature fluctuations and a full-f approach, including
consistent coupling of edge turbulence with scrape-off layer (SOL)
fluctuations and filamentary transport.

\section*{Acknowledgements}

We acknowledge main support by the Austrian Science Fund (FWF) project Y398.
This work has been carried out within the framework of the EUROfusion
Consortium and has received funding from the Euratom research and training
programme 2014-2018 under grant agreement No 633053. The views and opinions
expressed herein do not necessarily reflect those of the European Commission.  
The authors would like to thank F~C~Geisler, R~Kube and O~E~Garcia for valuable comments on the manuscript.


\section*{References}

\begin{thebibliography}{00}

\bibitem{bessenrodt93}
Bessenrodt-Weberpals~M \etal 1993 \NF {\bf 33} 1205

\bibitem{hawryluk98}
Hawryluk~R~J 1998 \RMP {\bf 70} 537

\bibitem{tokar04}
Tokar~M~Z, Kalupin~D and Unterberg~B 2004 \PRL {\bf 92} 215001

\bibitem{waltz04}
Waltz~R 2004 \PRL {\bf 93} 239501

\bibitem{tokar04b}
Tokar~M~Z 2004 \PRL {\bf 93} 239502

\bibitem{scott97} 
Scott~B~D 1997 \PPCF {\bf 39} 1635

\bibitem{scott92} 
Scott~B~D 1992 \PPCF {\bf 35} 1977 

\bibitem{dong94}
Dong~J~Q, Horton~W and Dorland~W 1997 {\it Phys. Plasmas} {\bf 1} 3635

\bibitem{lee97}
Lee~W~W and Santoro~R~A 1997 {\it Phys. Plasmas} {\bf 4} 169

\bibitem{xu13}
Xu~Y \etal 2013 \PRL {\bf 110} 265005

\bibitem{hahm13}
Hahm~T~S, Wang~L, Wang~W~X, Yoon~E~S and Duthoit~F~X 2013 \NF {\bf 53} 072002

\bibitem{pusztai11}
Pusztai~I, Candy~J and Gohill~P 2011 {\it Phys. Plasmas} {\bf 18} 122501

\bibitem{bustos15}
A. Bustos~A, Banon Navarro~A, G\"orler~T, Jenko~F and Hidalgo~C 2015 {\it Phys. Plasmas} {\bf 22} 012305

\bibitem{gurchenko16}
Gurchenko~A~D \etal 2016 \PPCF {\bf 58} 044002

\bibitem{scott03} 
Scott~B~D 2003 \PPCF {\bf 45} A385 
 
\bibitem{scott05b} 
Scott~B~D 2005 {\it Phys. Plasmas} {\bf 12} 102307

\bibitem{scott98} 
Scott~B~D 1998 {\it Phys. Plasmas} {\bf 5} 2334

\bibitem{scott01} 
Scott~B~D 2001 {\it Phys. Plasmas} {\bf 8} 447

\bibitem{scott07} 
Scott~B~D 2007 \PPCF {\bf 49} S25

\bibitem{arakawa66}
Arakawa~A 1966 {\it J. Comput. Phys.} {\bf 1} 119

\bibitem{karniadakis91}
G.E. Karniadakis~G~E, Israeli~M and Orszag~S~A 1991 {\it J. Comput. Phys.} {\bf 97} 414

\bibitem{naulin03}
Naulin~V and Nielsen~A 2003 {\it J. Sci. Comput.} {\bf 25} 104

\bibitem{kendl14}
Kendl~A 2014 {\it Int. J. Mass Spectrometry} {\bf 365/366} 106--113 

\bibitem{falchetto08}
Falchetto~G~L \etal 2008 \PPCF {\bf 50} 124015

\bibitem{madsen11}
Madsen~J \etal 2011 {\it Phys. Plasmas} {\bf 18} 112504

\bibitem{dorland93}
Dorland~W \etal 1993 \PF B {\bf 5.3} 812--835 

\bibitem{kendl06}
Kendl~A and Scott~B~D 2006 {\it Phys. Plasmas} {\bf 13} 012504

\bibitem{kendl03}
Kendl~A, Scott~B~D, Ball~R and Dewar~R~L 2003 {\it Phys. Plasmas} {\bf 10} 3684

\bibitem{hasegawa83} 
Hasegawa~A and Wakatani~M 1983 \PRL {\bf 50} 682 


\end{thebibliography}
\end{document}